\documentclass[amsmath,nofootinbib,twocolumn,showpacs]{revtex4-2}
\usepackage{float} % allows to use [H] for floats figures
\usepackage{enumitem} % allows to customize the enumerate function with different labels
\usepackage{subcaption} % allows subfigure
\usepackage{ragged2e}
\DeclareCaptionJustification{justified}{\justifying}
\usepackage{hyperref} % make references clickable links
\usepackage{graphicx}% Include figure files
\usepackage{endnotes}
\usepackage{hvfloat}
\usepackage{algorithm}
\usepackage{algcompatible}
\usepackage{booktabs}

\newcommand{\ra}[1]{\renewcommand{\arraystretch}{#1}}

\begin{document}
	\title{Agglomerative Likelihood Clustering}% Force line breaks with \\
	\author{Lionel Yelibi}
	\email{lionel.yelibi@alumni.uct.ac.za}
	\author{Tim Gebbie}
	\email{tim.gebbie@uct.ac.za}
	\affiliation{Department of Statistical Sciences, University of Cape Town, Rondebosch 7701, South Africa}
	\date{\today}

	\begin{abstract}
		We consider the problem of fast time-series data clustering. Building on previous work modeling the correlation-based Hamiltonian of spin variables we present an updated fast non-expensive Agglomerative Likelihood Clustering algorithm (ALC). The method replaces the optimized genetic algorithm based approach (f-SPC) with an agglomerative recursive merging framework inspired by previous work in Econophysics and Community Detection. The method is tested on noisy synthetic correlated time-series data-sets with built-in cluster structure to demonstrate that the algorithm produces meaningful non-trivial results. We apply it to time-series data-sets as large as 20,000 assets and we argue that ALC can reduce compute time costs and resource usage cost for large scale clustering for time-series applications while being serialized, and hence has no obvious parallelization requirement. The algorithm can be an effective choice for state-detection for online learning in a fast non-linear data environment because the algorithm requires no prior information about the number of clusters. 
	\end{abstract}
	\pacs{05.10.Ln, 75.10.Nr, 89.65.Gh}
	\keywords{Econophysics, Statistical Mechanics, statistics, Potts Models, Clustering, Phase Transitions, Time-series, finance}
	\maketitle
	
	% \tableofcontents 
	
	\section{Introduction} \label{sec:introduction}
	
	Summarizing large quantities of fast real-time feature time-series data that is sampled from environments with unknown dynamically evolving and non-linear interactions requires some sort of unsupervised learning\footnote{For a review on the last 20 years on financial market correlation based data clustering see \cite{marti_review_2017}, more generally on data clustering see \cite{jain_data_2010}, and particularly on time-series clustering see \cite{aghabozorgi_time-series_2015,liao_clustering_2005}.}. The problem of unsupervised statistical learning in the context of financial market data was explored in prior work \cite{yelibi_fast_2020} where the ability to quickly approximate super-paramagnetic cluster configurations \cite{blatt_superparamagnetic_1996,giada_data_2001} from data was shown. 
	Concretely, it was shown that the proposed algorithm recovered the correct super-paramagnetic cluster configurations near the entropy maxima. Previous cases studies include data clustering of stocks \cite{yelibi_fast_2020} and gene data in \cite{giada_algorithms_2002}, temporal states of financial markets \cite{marsili_dissecting_2002}, and state-detection for adaptive machine learning in trading \cite{hendricks_detecting_2016}. There is an endless variety of potential use-cases for this type of fast big-data clustering technology. The key problem with previous implementations related to the compute times for very large data-sets.
	
	Building on prior work we propose and demonstrate an alternative to fast Super-Paramagnetic Clustering (f-SPC) \cite{yelibi_fast_2020} using a modern and streamlined implementation of the ``Merging Algorithm" first suggested in \cite{giada_algorithms_2002}, one that can recover the same or better cluster configurations but with significantly reduced compute times. The algorithm is in spirit similar to network science community detection algorithms \cite{ronhovde_local_2010,blondel_fast_2008}.
	
	The analysis of structure of stock market correlation matrices constitute a large body of study in econophysics, quantitative finance, and network science. Representing the relationships between financial assets in terms of networks has become an increasingly useful approach to explore, gain insights, and visualize the collective impact of various aggregate dynamical process on assets; be it the impact of a market crashes, or that of the iterative application of a portfolio optimization. This is convenient because graphs can be represented by adjacency matrices, and correlation matrices can then be mapped to network data before they are processed; here passed through a community detection algorithm. 
	
	Network data is usually represented by sparse matrices, this can operationally be seen as the main difference between data clustering, and community detection -- it is the difference in representation. In order to map correlation matrices into adjacency matrices a threshold is often applied \cite{tumminello_correlation_2010,namaki_network_2011,zalesky_use_2012}. The threshold value above which correlated nodes are linked or unlinked is arbitrary. However, this can add a layer of complexity in the study of graphs of financial assets, while also being an additional source of noise that has little to do with the dynamic processes governing the relationships between assets, or objects more generally. Our approach differs from threshold based approaches in that it is a faithful data clustering model which works with dense correlation matrices -- this can make it appropriate for complex systems in noisy environments but without the arbitrariness of threshold choice.
	
	Here, we again exploit the Noh Ansatz \cite{noh_model_2000} and the Maximum Likelihood Estimation (MLE) approach first introduced by Giada and Marsili \cite{giada_data_2001,giada_algorithms_2002}. We call the new algorithm Agglomerative Likelihood Clustering (ALC); it has the benefit of being less computationally expensive than the Parallelized Genetic Algorithms (PGAs) previously implemented in \cite{yelibi_fast_2020,hendricks_detecting_2016,hendricks_high-speed_2016}. The performance enhancement arises from being able serialize the algorithm into a brute-force search across cluster configurations so as to avoid unnecessary computational expense of genetic algorithm used in, {\it e.g.} the f-SPC implementation. 
	
	The paper is as follows: in Section \eqref{sec:giada_marsili} we introduce the Giada-Marsili model, in Section \eqref{sec:alc} we describe the optimization algorithm, in Section \eqref{sec:synthetic} we consider clustering of synthetic generated time-series data, Section \eqref{sec:bootstrap} provides additional  tools for when the signal-to-noise ratio is relatively low, Section \eqref{sec:nested} explores the solutions recovered for complex correlation matrices with nested block structure, Section \eqref{sec:performance} discusses and compares our new method execution runtime to the previous algorithm and HDBSCAN,this is followed by the discussion and conclusion in Section \eqref{sec:discussion} that highlights the performance of the algorithm we have introduced. 
	
	\section{Giada-Marsili Likelihood Model} \label{sec:giada_marsili}
	
	The spectral analysis of stock market correlation matrices provides an intuition taking the form of the { \it Noh ansatz} \cite{noh_model_2000}: that there exists a hierarchical structure in financial markets where individuals stocks are sub-components of larger groups of assets, and each asset's individual returns is influenced by the collection of assets it belongs to. This can be expressed in the form of a simple generative model \cite{noh_model_2000}:
	\begin{equation} \label{eq:7} 
	x_i = f_i + \epsilon_i 
	\end{equation} 
	where $x_i$ are the stock's features, $f_i$ the cluster-related influence, and $\epsilon_i$ the node's specific effect.
	
	In turn, this generative model can be iteratively nested within a more complex set of independent hierarchies as expanded out through the unexplained noise terms in the presence of top-down and bottom-up causal relationships \cite{wilcox_hierarchical_2014,tumminello_hierarchically_2007}. Here we restrict ourselves to a single collection that is broken up into groups with a simple noise terms for unexplained externalities.
	
	We consider a group of $N$ observations embedded in a space with dimensionality $D$ as the features, every observation is assigned a spin value. One version of the ansatz models the observation features such that \begin{equation} \label{eq:ansatz_1} x_i = g_{s_i}\eta_{s_i} + \sqrt{1- g_{s_i}^2}\epsilon_i \end{equation} where $x_i$ is one feature, $g_{s_i}$ the intra-cluster coupling parameter \footnote{The thermal average $\langle g_s \rangle$ can be used to reconstruct data-sets sharing identical statistical features of the original time-series by using Eqn. \eqref{eq:ansatz_1} \cite{giada_data_2001}}, $\eta_{s_i}$ the cluster-related influence, and $\epsilon_i$ the observation's specific effect and measurement error. A covariance analysis yields additional terms such as $n_s$ the size of cluster $s$, and $c_s$ the intra-cluster correlation \footnote{Here $n_s = \sum_{i=1}^{N} \delta_{s_i,s}$, $c_s = \sum_{i=1}^{N}\sum_{j=1}^{N}C_{ij}\delta_{s_i,s}\delta_{s_j,s}$, and $g_s = \sqrt{ \frac{c_s-n_s}{n_s^2-n_s}}$ \cite{giada_data_2001,hendricks_detecting_2016}.}.
	
	We explicitly mention that $n_s < c_s < n_s^2$ must be enforced: the lower bound is required because $g_s$ is undefined for values of $c_s \leq n_s$, and the upper bound requires a strict inequality because Eqn. \eqref{eq:lc} is undefined when $c_s = n_s^2$. We introduce a Dirac delta function \footnote{Let $ y_i = x_i - \big(g_{s_i}\eta_{s_i} + \sqrt{1- g_{s_i}^2}\epsilon_i\big)$, and $\delta(y)$ a Dirac delta function of $y$ which is 1 when $y=0$, and 0 otherwise.} to model the probability of observing data in a configuration $S$ close to criticality\cite{mastromatteo_criticality_2011}:
	\begin{equation} \label{eq:12} 
	P = \prod_{d=1}^{D}\prod_{i=1}^{N}\Bigg \langle \delta(x_i - ( g_{s_i}\eta_{s_i} + \sqrt{1 - g_{s_i}^2}\epsilon_i)) \Bigg \rangle~.
	\end{equation}
	This joint likelihood is the probability of a cluster configuration matching the observed data for every observations, and for every feature. The log-likelihood derived from $P$ can be thought of the Hamiltonian of this Potts system \cite{wu_potts_1982}: 
	\begin{equation} \label{eq:lc} 
	L_c = \frac{1}{2} \sum_{s:n_s>1} \bigg[ \ln \frac{n_s}{c_s} + (n_s-1) \ln \frac{n_s^2-n_s}{n_s^2-c_s} \bigg] ~.
	\end{equation}
	The sum is computed for every feature, and represents the amount of structure present in the data. The value of $L_c$ is indirectly dependent on spins via the terms $n_s$ and $c_s$. 
	
	There are advantages to this method relative to most industry standard alternatives: First, that $L_c$ is completely dependent on $C_{ij}$, and the dimensionality of the data-set only plays a part in computing $C_{ij}$. Second, it is adaptive: the number of clusters is not given as an input, unlike K-MEANS \cite{steinley_k-means_2006} or similar algorithms. Clustering configurations are randomly generated, and that which maximizes $L_c$ provides us with the number of clusters, and their compositions. The importance of this later feature should not be lost as it removes a layer of unnecessary hyper-parameter tuning. As such the algorithm competes with the likes of DBSCAN \cite{ester_density-based_1996,khan_dbscan_2014} and HDBSCAN \cite{campello_density-based_2013,mcinnes_hdbscan_2017}.
	
	\section{Agglomerative Likelihood Clustering} \label{sec:alc}
	
	A traditional perspective often use when considering clustering problems is to try and demarcate methods into either those that implement either top-down, or bottom-up algorithms. Using this perspective, top-down methods are then thought of as divisive and consist in starting with a single cluster as initial condition and splitting (or partitioning) the graph in additional clusters iteratively while minimizing the cost. On the other hand, bottom-up methods initially start with each observations in their own clusters, and proceed to merge them iteratively \cite{murtagh_algorithms_2012}. 
	
	The so called ``Louvain" algorithm \cite{blondel_fast_2008} is agglomerative and can then be argued to implement the later bottom-up approach to ``community detection" on networks. It is in spirit very similar to the Merging Algorithm (MR) developed by Marsili and Giada in \cite{giada_algorithms_2002}.
	
	Our previously implemented methods were based broadly on efficient implementations that streamlined PGA optimization frameworks, {\it e.g.} those proposed in \cite{yelibi_fast_2020,hendricks_detecting_2016,hendricks_high-speed_2016} allow for all sorts of mutations. However, these approaches can be sensitive to initial conditions because at every step a new generation of individuals is mutated, evaluated, and then a group of the best candidates survives until the next algorithm's iteration  -- this can be arbitrarily path dependent. The f-SPC algorithm \cite{yelibi_fast_2020} secured additional computational advantages within this PGA framework by being able to exclude a computationally expensive mutation constraint within the genetic programming framework used in \cite{hendricks_detecting_2016,hendricks_high-speed_2016}.
	
	However, genetic algorithm based approaches have disadvantages, the key ones are discussed in Table \eqref{tab:mle_dis} and relate to the need for : i.) ambiguous stopping or convergence criteria, ii.) random mutations, and iii.) some sort of parallelization to reduce compute times. 
	
	\begin{table}[H]
		\noindent\fbox{\parbox{.45\textwidth}{
				\begin{enumerate}[label=\eqref{tab:mle_dis}\arabic*]
					\item {\bf Convergence Criteria}: Assuming the existence of multiple local maxima it tries to navigate around these ``sub-optimal" solutions on its way to a global maximum. However, there is no certainty and it is just assumed that the algorithm stops once a criteria is met - the algorithm is explicitly stochastic in convergence.
					\item {\bf Random Mutations}: Because the algorithm applies random mutations, the population size, the number and diversity of mutations, and the number of generations all have an impact on the final result - this can introduce path dependence. 
					\item {\bf Parallelization}: The algorithm requires evaluating the entire mutated population at every iteration. This has both a computational and memory cost as it requires loading the data (i.e. the correlation or similarity matrix) on every worker. The Likelihood evaluation itself is inexpensive but multi-processing adds CPU-overhead. This can be mitigated by using GPUs as in \cite{hendricks_detecting_2016}.
		\end{enumerate}}}
		\caption{\label{tab:mle_dis} Disadvantages of the PGA algorithms for the likelihood $L_c$. }
	\end{table}
	
	\subsection{Greedy Merging} \label{ssec:greedy}
	
	To build a fast general bottom-up merging algorithm we again start with all $N$ spins in $N$ cluster, but we iteratively merge clusters in a greedy fashion.
	
	The Giada-Marsili Merging Algorithm (MR) implementation requires computing the change in likelihood $\Delta L_c$: We consider three clusters $C_1$, $C_2$, and $C_3$ with $C_3=C_1 + C_2$ where the addition operator ``+'' means clusters $C_1$ and $C_2$ are merged. Marsili and Giada define two cases for $\Delta L_c$ \cite{giada_algorithms_2002}:
	\begin{eqnarray}
	%\centering
	&\mathbf{Case~1:}&~\Delta L_c = L_c(C_3) - \max [ L_c(C_1), L_c(C_2) ]\label{eq:dellcmax} \\
	%\centering
	&\mathbf{Case~2:}&~\Delta L_c = L_c(C_3) - [ L_c(C_1) + L_c(C_2) ]\label{eq:dellc}
	\end{eqnarray}
	In Case 1, as described by Eq. \eqref{eq:dellcmax}, $C_3$ would be a better cluster than any of $C_1$ and $C_2$. Here we chose to use the more restrictive definition, Case 2, as defined by Eq. \eqref{eq:dellc}. The key insight is to realize that Case 2 requires that the new merged cluster must be better than the combination of the two individual sub-clusters. We can iteratively exploit this by building an algorithm that performs a comprehensive grid search over the space of clusters.
	
	To implement this we can modify Eq. \eqref{eq:lc} by removing the sum and only compute the likelihood of individual clusters: 
	\begin{equation} \label{eq:lcmod} L_c = \frac{1}{2} \bigg[ \ln \frac{n_s}{c_s} + (n_s-1) \ln \frac{n_s^2-n_s}{n_s^2-c_s} \bigg] ~.
	\end{equation}
	The objective at every iteration is to then maximize $\Delta L_c$ over every possible move.
	\text
	The implementation we adopt to generate the moves is inspired by innovations made in community detection methods, {\it i.e.} community detection \cite{javed_community_2018} algorithms such as the ``Louvain algorithm" \cite{blondel_fast_2008}. 
	
	Using this type of community detection we return to a bottom-up agglomerative approach to quickly enumerate candidate configurations using a likelihood method to determine the efficacy of the particular configuration. The agglomerative likelihood clustering framework is a general framework suited for any useful choice of likelihood function. 
	
	Here we will make the choice of the modified Giada-Marsili likelihood function, and then using Case 2 from the Merging Algorithm (Eqn. \eqref{eq:dellc}). There are other choices of likelihood that could be made {\it e.g} a multi-factor cluster specification, or the likelihood function for K-means. The framework is general - but the choice made here is specific to the spin Hamiltonian framework. 
	
	\subsection{Iterating cluster configurations} \label{ssec:tracker}
	
	The implementation requires keeping track of both the correlation matrix, and the cluster configurations at each iteration of the algorithm. In order to retain a flexible but dynamic representation of the cluster configurations we introduce the idea of the tracker array (see Table \eqref{tab:ALC_init}). This object will store the list of objects represented in each cluster during algorithm updates\footnote{In {\tt python} the {\tt numpy} computed correlation matrix can be stored to a python dictionary so that it can be easily and efficiently modified. The tracker can be stored as a list variable.}. The algorithm is initialized with singleton clusters.
	
	Once the initialization step in Table \eqref{tab:ALC_init} is performed, we can move onto the actual optimization steps required, these are described in Table \eqref{tab:ALC}. This requires a cluster update step, computing the likelihood changes, updating the record of configurations using that tracker lists, updating the correlation matrix and then checking the convergence criteria.
	
	\begin{table}[H]
		\noindent\fbox{\parbox{.45\textwidth}{
				\begin{enumerate}[label=\eqref{tab:ALC_init}\arabic*]
					\item {\bf The Correlation Matrix:} $C$  is stored in a structure that can be easily modifiable whenever new entries are needed, additional clusters created, or when previously clustered objects are removed.
					\item {\bf Singleton Initialization}: All objects start in their own clusters.
					\item {\bf Tracker List}: We create a list which stores lists of objects: each list represents a cluster, and the labels inside the list are cluster members. The variable $n_s$ is the cluster size.	
		\end{enumerate} }}
		\caption{The ALC initialization uses a singleton configuration, a correlation matrix stored in an efficient manner that can be easily modified during the algorithm implementation, and a tracker list that records the configurations.}
		\label{tab:ALC_init}
	\end{table}  
	
	\begin{table}[H]
		\noindent\fbox{\parbox{.45\textwidth}{
				\begin{enumerate}[label=\eqref{tab:ALC}\arabic*]
					\item {\bf Clustering}: Pick an object at random in the tracker's labels, cluster it with all other objects, and store the resulting $\Delta L_c$ values.
					\item {\bf Maximize Likelihood Change}: Find out the highest $\Delta L_c$; and then if it is bigger than 0 continue to the next step. Else the object cannot be clustered, is removed from the list  and the process restarts.
					\item  {\bf Tracker Update}: if there exists a positive $\Delta L_c$, create a new label for the new cluster and its content is the union of the two clusters we merge.
					\item  {\bf Correlation Matrix Update}: The correlation values are the sums of the correlations of the two clusters merged. The self-correlation is the sum of the intra-cluster correlations.
					\item  {\bf Iterative Convergence}: This process is repeated until $\Delta L_c$ is non-negative.
		\end{enumerate} }}
		\caption{The main routine of agglomerative likelihood clustering algorithm for time-series clustering consists of merging clusters to find a maximum increase in likelihood (See pseudo-code in Appendix \eqref{alg:alc})}
		\label{tab:ALC}
	\end{table}

	\begin{figure*}
		\begin{subfigure}{.3\textwidth}
			\includegraphics[width=\textwidth]{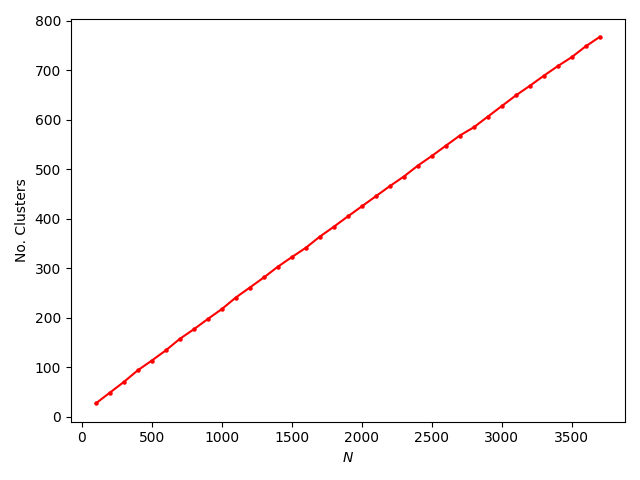}	
			\caption{Average Number of clusters per clustering solution as a function of Data-set Size.}
			\label{fig:nclnoise}
		\end{subfigure}	
		\begin{subfigure}{0.3\textwidth}
			\includegraphics[width=\textwidth]{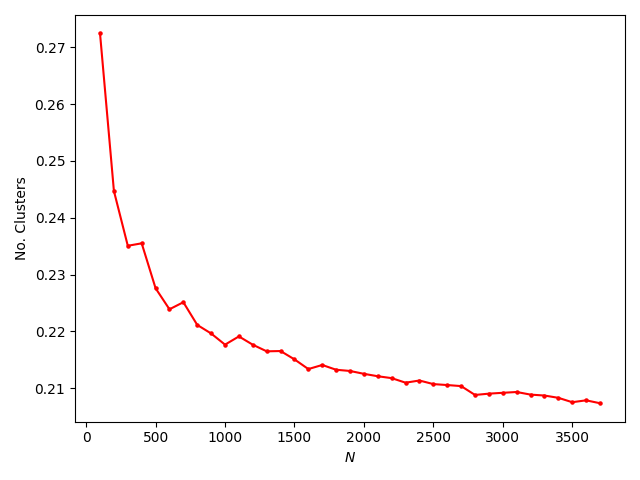}
			\caption{Normalized Number of Cluster with respect to the data-set Size as function of the data-set Size.}
			\label{fig:nclnormnoise}
		\end{subfigure}
		\begin{subfigure}{0.3\textwidth}
			\includegraphics[width=\textwidth]{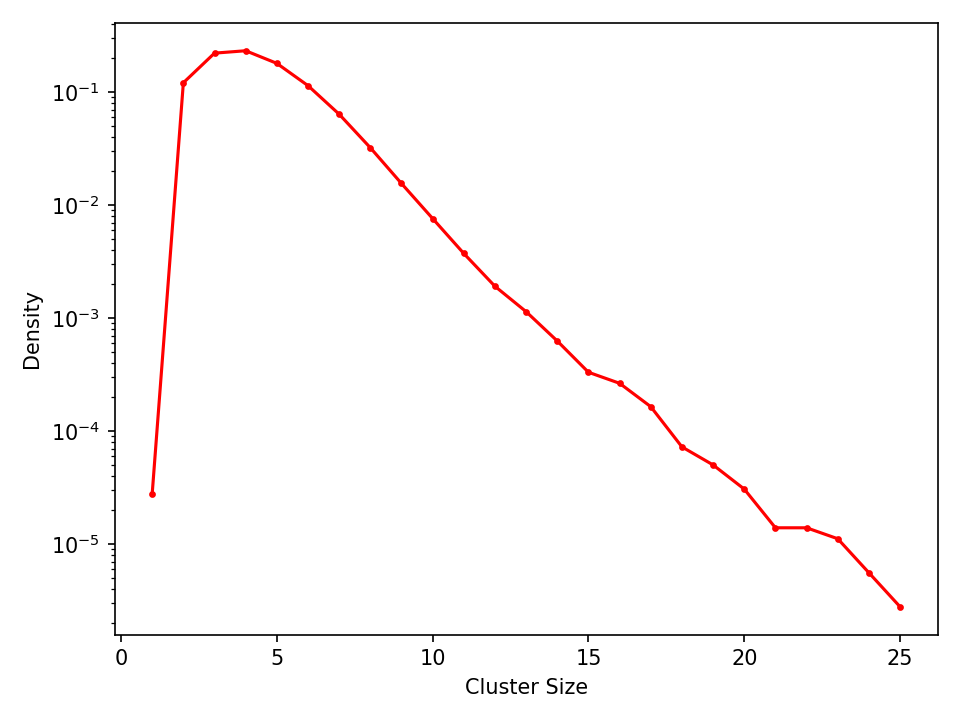}
			\caption{Semi-log plot of the distribution of cluster size for uncorrelated cluster-less data-sets.}
			\label{fig:noisedist}
		\end{subfigure}
		\caption{ Descriptive Analysis of data-clustering of uncorrelated clusterless time-series data.} \label{fig:}
	\end{figure*}
	
	\begin{figure*}
		\begin{subfigure}{\textwidth}
			\includegraphics[width=\textwidth]{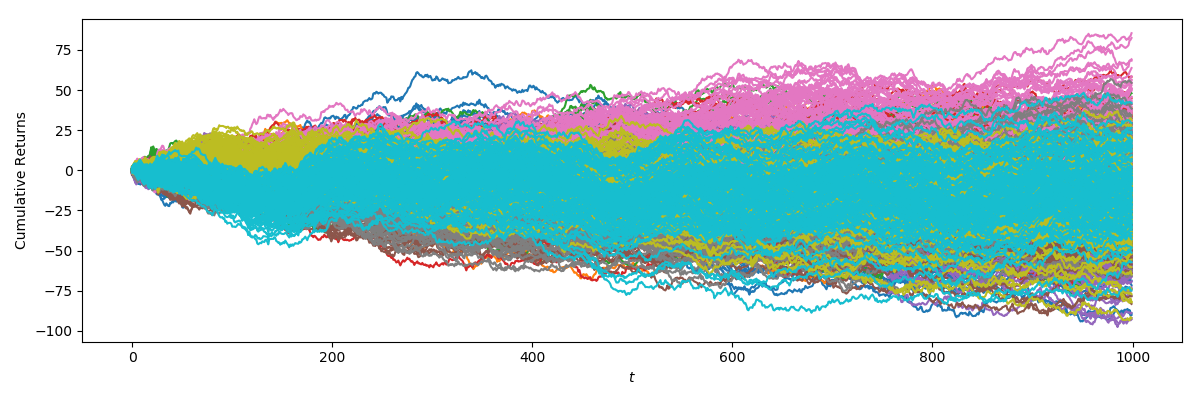}
			\caption{{\bf Simulated}: Cluster derived correlated time-series cumulative returns for 500 simulated assets over 1000 days.}
			\label{fig:correlated_timeseries}
		\end{subfigure}
		\begin{subfigure}{0.47\textwidth}
			\includegraphics[width=\textwidth]{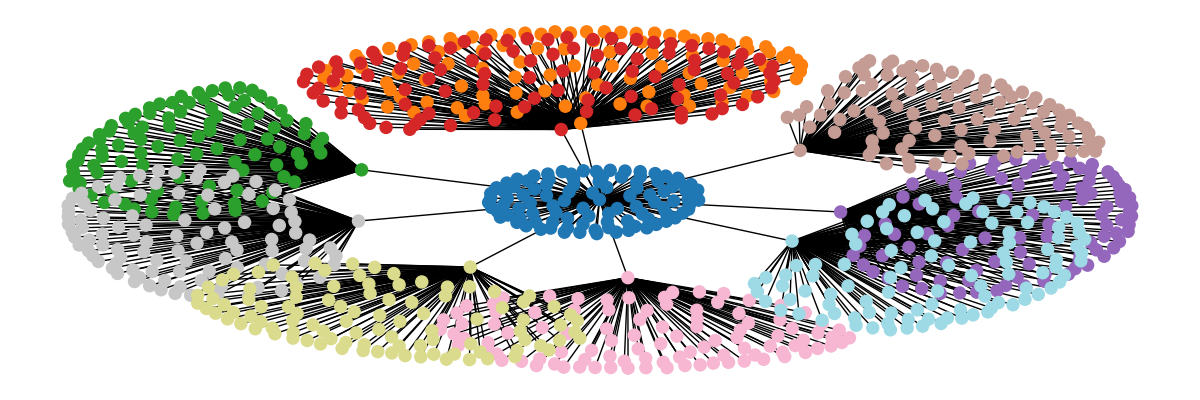}
			\caption{{\bf Ground Truth}: The true correlation matrix MST }
			\label{fig:mst_gm}
		\end{subfigure}
		~
		\begin{subfigure}{0.47\textwidth}
			\includegraphics[width=\textwidth]{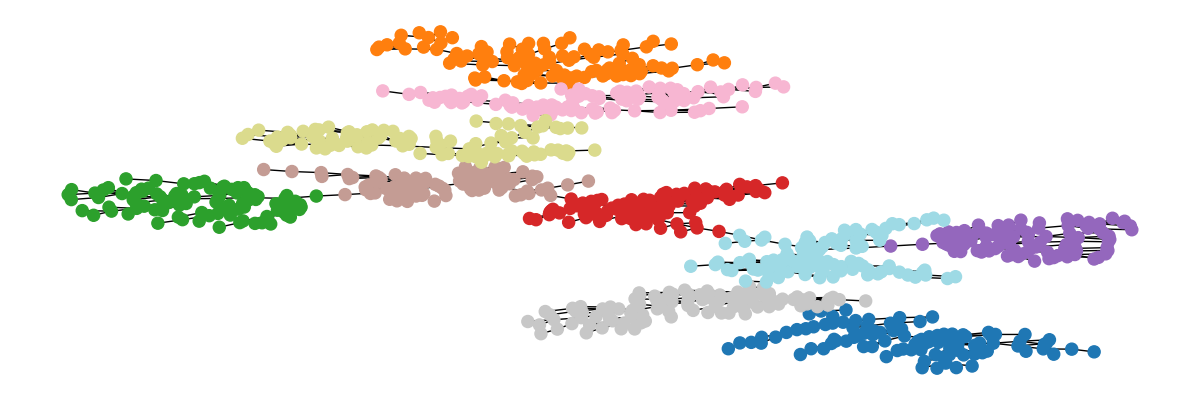}
			\caption{{\bf Estimated}: The estimated correlation matrix MST }
			\label{fig:mst_est}
		\end{subfigure}
		\caption{Using synthetic correlated time-series data created with Table \eqref{tab:gen_timeseries}. In subplot a.) 500 Normalized time-series cumulative daily returns colored by cluster. And respectively in b.) and c.) The Minimum Spanning Trees of the true cluster configuration, and the estimated correlation matrix from the synthetic data. The colours represent the 10 clusters present in the data.}
	\end{figure*}
	
	\section{Synthetic Data} \label{sec:synthetic}
	
	\subsection{Noise and uncorrelated data} \label{ssec:synthetic_noise}
	
	First, we start testing our algorithm on random uncorrelated data ``white noise''. We generate uncorrelated data from the student-t distribution with 3 degrees of freedom with different data-set sizes and lengths. in Figure \eqref{fig:noisedist} we can observe the probability distribution of cluster sizes recovered by the algorithm. Uncorrelated data yields clusters despite the non-existence of clusters by design. Additionally in Fig. \eqref{fig:nclnoise} we show the number of ``noise'' clusters with increasing data-set size: The linear relationship between number of clusters and data-set size is trivial, however when the cluster number is normalized with $N$ we observe that it in fact grows slower than $N$. The curve in Fig. \eqref{fig:nclnormnoise} could serve as the upper bound on our method's solutions. 
	
	The problem of identifying clusters when they are in fact not present in the data can be explained by estimation noise inherent to sample covariance estimation. The ground truth is the identity matrix, but due to finite size effects the estimated off-diagonal values are non-zero. The clusters sizes appear fat tailed on a semi-log plot, and the mode of the distribution is 5. 
	
	The implications are multiple: Small clusters of size 5 and below could be statistically indistinguishable from noise, while larger clusters are less likely to be spurious. Furthermore, true uncorrelated data should yield solutions of singletons which is not absolutely the case because as previously mentioned the number of ``noise'' doesn't grow as fast as $N$.	
	
	\subsection{Clustering Synthetic Student-t data} \label{ssec:synthetic_correlated}
	% for various sizes and lengths
	
	\begin{table*}\centering
		\ra{1.3}
		\begin{tabular}{@{}lrrrcrrr@{}}\toprule
			& \multicolumn{2}{c}{ALC} & \phantom{abc}& \multicolumn{3}{c}{HDBSCAN}\\
			\cmidrule{2-4} \cmidrule{6-8}
			& $N=500$ & $N=1500$ & $N=3000$ && $N=500$ & $N=1500$ & $N=3000$\\ \midrule
			$g_s=0.05$\\
			$D=20$ & 0.04 & 0.06 & 0.03 && 0.03 & 0.04 & 0.00\\
			$D=60$ & 0.07& 0.07& 0.07&& 0.03& 0.03& 0.02\\
			$D=250$ & 0.19& 0.14& 0.14&& 0.09& 0.08& 0.09\\
			$g_s=0.1$\\
			$D=20$ & 0.09& 0.08& 0.11&& 0.04& 0.01& 0.04\\
			$D=60$ & 0.19& 0.15& 0.14&& 0.12& 0.07& 0.03\\
			$D=250$ & 0.47& 0.39& 0.32&& 0.23& 0.18& 0.12\\
			$g_s=0.3$\\
			$D=20$ & 0.27& 0.24& 0.24&& 0.22& 0.17& 0.12\\
			$D=60$ & 0.58& 0.49& 0.41&& 0.40& 0.31& 0.20\\
			$D=250$ & 0.90& 0.84& 0.76&& 0.66& 0.60& 0.35\\
			$g_s=1$\\
			$D=20$ & 0.61& 0.49& 0.44&& 0.56& 0.51& 0.48\\
			$D=60$ & 0.90& 0.87& 0.81&& 0.83& 0.80& 0.73\\
			$D=250$ & 0.99& 0.98& 0.96&& 0.98& 0.94& 0.92\\
			\bottomrule
		\end{tabular}
		\caption{The Adjusted Rand Index (ARI) for the ALC and HDBSCAN cluster solutions for data-sets containing 10 clusters of correlated time-series. From left to right the size varies from 500 to 3000 while from top to bottom time-series length goes from 20 to 250. $g_s$, the average intra-cluster correlation, is also changed from 0.05 to 1.} \label{tab:synth}
	\end{table*}
	
	Complex systems are characterized by their multi-scale dynamics in time and space. Financial markets are a good example of this where stock prices are recorded and aggregated on time-scales ranging from milliseconds to daily prices, if not weeks, months and years. There can be vastly different if not incompatible data generating processes operating on different time-scales. At the same time, markets are intrinsically non-stationary, which means data going far back in the past becomes stale, and this has a clear impact on most investment models performance.
	
	Here we consider correlated synthetic data and generate Student-t distributed data-sets of 500, 1500, and 3000 time-series of variables lengths 60, 250, and 500 using the one-factor model in Eq. \eqref{eq:ansatz_2} (See Appendix \eqref{addix:onefactor}). Gaussian mixtures are readily used within the machine learning literature to test data clustering. Here we simply use the student-t distribution because stock returns distributions are known to be fat-tailed \cite{gabaix_theory_2003,gopikrishnan_inverse_1998}. This stylized fact is some what captured by simulating with the student-t distribution; we would like to investigate the potential effect this may have on our algorithm. 
	
	The data-sets of sizes: 500, 1500, and 3000 seem arbitrarily chosen. However, the size of the S\&P500 is roughly 500, and Indices of the Russell family: Russell 1000, 2000, and 3000 track up to 90\% of the US stock market total market capitalization. Large portfolio of thousands of stocks exist for which equivalently large correlation matrices must be estimated. Correlation matrices are readily directly used in Markowitz's portfolio optimization \cite{zhang_portfolio_2018,kalayci_comprehensive_2019} or indirectly via clustering using methods such as hierarchical risk parity \cite{prado_building_2016}.
	
	The intra-cluster coupling strengths $g_s$, were arbitrarily selected to illustrate two limiting cases: 0.05 and 1, and two intermediate cases: 0.1, and 0.3. Here the value 0.3 is an approximation to the average stock market average correlation \cite{pollet_average_2010}. This is not the average intra-cluster correlation, which is what $g_s$ is; nevertheless, it gives a sense of how the algorithm would perform given real stock market data. The limiting values 0.05 and 1 serve to show how performance degraded ($g_s=0.05$) when dealing with low or weakly bound clusters, as opposed to perfectly correlated ones ($g_s=1$). Given two time-series belong to different clusters, their pairwise correlation should be 0. However, this is not always the case because there is a noise-floor, and a lower $g_s$ will bring the clusters close to the noise floor. We have empirically confirmed in Table \ref{tab:synth} that this will deteriorate the algorithms performance. Here each of the size and length pairs was simulated a hundred times, and the Adjusted Rand Index (ARI) \cite{santos_use_2009} between the ground truth and the model's output was averaged (See Table \eqref{tab:synth}).
	
	The time-series length were chosen with the noise floor effects in mind. Short time-series are expected to induce a higher noise floor, and spurious correlations. While longer time-series ameliorate this even though finite size effects remain. Towards this end we select the lengths: 20, 60, and 250. These are are approximations for the number of trading days in 1, 3, and 12 months of real world data. These lengths are adequate for data clustering on daily prices. It also follows that within a high frequency regime, one where measurements are abundant, correlation matrices are easier to estimate, and the algorithm should perform accordingly. However, in the high-frequency domain one is more directly concerned about the applicability of sample estimated correlations that do not deal with the impact of discretization \cite{pedregosa_scikit-learn_2011}.
	
	\section{Noise: Re-sampling Clusters and Bootstrap} \label{sec:bootstrap}
	%can mitigate issues with low dimensional data.
	
	In Sec. \eqref{ssec:synthetic_correlated} we show that the estimation of correlation matrices induces statistical noise which significantly affects the performance of clustering algorithms. Here we provide a way to mediate the influence of noise using bootstrapping. The intuition arises from the relationship between the algorithm performance with respect to the signal to noise ratio $Q = \frac{D}{N}$. 
	The analysis of correlation matrices using Random Matrix Theory shows that they can be exactly estimated in the limit where $N \rightarrow \infty$, $D \rightarrow \infty$ and $Q \geq 1$ and is maintained. 
	
	Despite the illusion of living in the so-called era of ``Big Data'', in Finance, and more specifically in the context of financial markets, observables remain non-stationary and subject to extreme events, shocks and regime changes, all in the presence of strategic purposeful agents. Data quickly becomes obsolete. This means that the length of time-series used to estimate the correlation matrix $C_{ij}$ can be pragmatically shortened, or adaptively estimated, to better capture more relevant, time or regime dependent recent dynamics, but at the cost of increasing estimation noise. As shown in Table \eqref{tab:synth} this can have a severe impact on clustering solutions.
	
	\subsection{Filtered correlation matrix} \label{sec:filtered}
	
	Here we consider cases in which the number of stocks $N$ is large, but the number of realisations $D$ is small, leading to $Q \leq 1$ to violate the quality conditions. We then propose that a definitive filtered cluster membership matrix be constructed by sampling $n$ in $N$ stocks such that the quality condition on the $n$-sampled sub-problem is satisfied: $q = \frac{D}{n} \geq Q$. The routine that implements this  is made explicit in Table \eqref{tab:boot}. This introduces the idea of a filtered correlation matrix $c_{ij}$, which is inspired from the original Potts ``spin-spin correlation function'' \cite{wu_potts_1982}. 
	
	Consider ${\cal{M}}$ random draws of $n$ objects from a collection $\{1,...,N\}$. First, from each sample we computed the pair-wise frequency of objects being drawn together, $f_{ij}$:
	\begin{equation}
	f_{ij} = \sum_{m \in \cal{M}} \sum_{i < j} \delta_{ij} \label{eq:boot_sample_freq}.
	\end{equation}
	Second, we find the pair-wise frequency of objects begin clustered together, $d_{ij}$:
	\begin{equation}
	d_{ij} = \sum_{m \in \cal{M}} \sum_{i < j} \delta_{s_i s_j}. \label{eq:boot_sample_clustered} 
	\end{equation}
	These can then be combined into a pair-wise normalised probability of clustered objects: $p_{ij}$:
	\begin{equation}
	p_{ij} = \frac{d_{ij}}{f_{ij}}.\label{eq:boot_filtered_correlation}
	\end{equation}
	This is a measure of the probability that objects will being clustered together. This is then mapped to an ordinal codependency function on $[0,1]$ using a threshold $\omega$:
	\begin{equation}
	p^{\omega}_{ij} = \begin{cases}
	1 ,& p_{ij}-\omega > 0,\\
	0  , & p_{ij}-\omega \le 0. 
	\end{cases} \label{eq:boot_ordinal}
	\end{equation}
	The final step is similar to the thresholding method used in the original Super-Paramagnetic Clustering  (SPC) algorithm proposed by \citet{blatt_data_1997}\footnote{The original SPC uses simulated annealing, which via a repeated pairwise linkage validation process estimates a quantity similar to $p_{ij}$ (See. Eq. \eqref{eq:boot_filtered_correlation}) called the ``spin-spin correlation''}. We note that any activation function could be used, and it is reminiscent of a logistics regression. 
	
	\begin{figure*}
		\begin{subfigure}{.45\textwidth}
			\includegraphics[width=\textwidth]{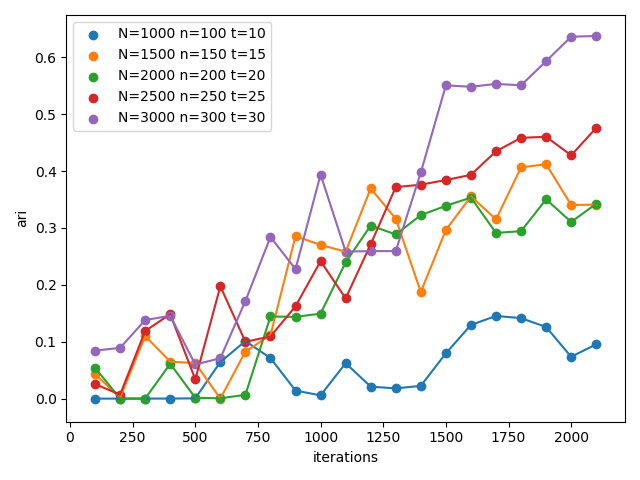}	
			
			\caption{ARI for Bootstrap ALC using the threshold $\omega = 0.5$.}
			\label{fig:surr05}
		\end{subfigure}	
		\begin{subfigure}{0.46\textwidth}
			\includegraphics[width=\textwidth]{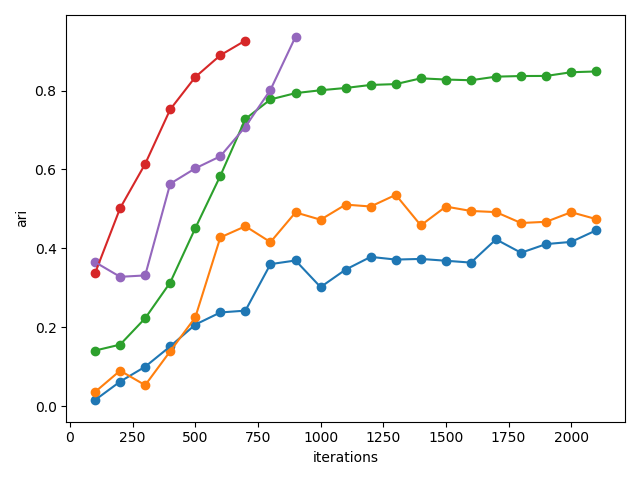}
			\caption{ARI for Bootstrap ALC using the threshold $\omega = 0.75$. }
			\label{fig:surr075}
		\end{subfigure}
		\caption{The ARI is computed for the cluster solutions recovered using the bootstrap method and the ground truth. Data-sets size varies from 1000 to 3000, and the Signal-to-Noise ratio is fixed at $q=0.1$ } \label{fig:surr}
	\end{figure*}
	
	\begin{table}[H]
		\noindent\fbox{\parbox{.45\textwidth}{
				\begin{enumerate}[label=\eqref{tab:boot}\arabic*]
					\item {\bf Hyper-parameters} ($q$,$n$,$\omega$,$m$): 
					\subitem Target signal-to-noise ratio: $q$. \subitem Sample size $n$ computed from $q$. 
					\subitem Correlation threshold: $\omega$.
					\subitem Number of iterations: $m$.
					
					\item {\bf Resample Clusters: $k=1$ to $m$}
					
					\subitem 1. Compute $N \times N$ matrix $f^{(k)}_{ij}$,
					\subitem 2. Cluster sample of $n$ indices, 
					\subitem 3. Compute $N \times N$ matrix $d^{(k)}_{ij}$.
					
					\item  {\bf Probability matrix }: Compute the
					$N \times N$ matrix $p_{ij}$ (Eqn. \eqref{eq:boot_filtered_correlation}).
					% this is a component wise operation?
					\item  {\bf Ordinal Filtered Dependency matrix}: Compute $p^{\omega}_{ij}$ (Eqn. \eqref{eq:boot_ordinal})
					\item  {\bf Final Configuration}: The final cluster configuration can then be built from the ordinal matrix $p^{\omega}_{ij}$ acting as the adjacency matrix of a graph.
		\end{enumerate}}}
		\caption{The bootstrapping routine for Agglomerative Likelihood Clustering (See Python script ``cluster\_resampling.py'' \cite{yelibi_agglomerative_2020}). The goal is to repeatedly cluster subsets of the data-set aggregated in $d_{ij}$ (Eq. \eqref{eq:boot_sample_clustered}), transformed in normalized probability $p_{ij}$ (Eq. \eqref{eq:boot_filtered_correlation}) and filtered into ordinal filtered dependency $p^{\omega}_{ij}$ (Eq. \eqref{eq:boot_ordinal}) from which the final cluster is extracted.} \label{tab:boot}
	\end{table}      
	
	The $n$ indices sampled are included in $N$, this means that the routine will cluster small subsets of the data-set for which the correlation matrix has a higher quality factor $q$ than $Q$. The operation is repeated for a preset number of iterations $m$, and we use both matrices A and F to create the $C^*$. The threshold function is then applied to $C^*$ to create a filtered adjacency matrix with disconnected components from is extracted the cluster solution: $C^{F*}$.
	
	The number of maximum iterations $m$ is arbitrarily chosen, in experiments we show that the convergence of the algorithm depends on $n$ and indirectly on $N$. This means that larger data-sets require more sampling of the index space. We arbitrarily set the maximum iteration number: $ m= 2200$.	
	
	Finally, we explored two threshold values: $\omega = 0.5$ and $\omega=0.75$ and we compare this to using the niave correlation matrix. Although arbitrary, the intuition comes from \cite{blatt_data_1997} where the threshold used to cluster objects in the same group is a half: $0.5$. 
	
	The bootstrap method was tested on several correlated data-sets whose sizes and time-series length ranged from $N=1000$ and $D=10$ to $N=3000$ and $D=30$. The $n$ sample size was chosen such that $q=0.1$ given $Q=0.01$. The output cluster solutions were compared against the ground truth using the ARI. The ARI is also used as a stopping condition: if $\mbox{ARI} \geq 0.9$ the algorithm stops.
	
	Figures \eqref{fig:surr05} and \eqref{fig:surr075} show thresholding with 0.5 is significantly worse than 0.75. In Fig. \eqref{fig:surr05} the solutions have yet to converge after 2000 iterations. Although they seem to be following an upward trend all trajectories are absolutely inferior to their Fig. \eqref{fig:surr075} counterparts after 500 iterations.
	
	In Fig. \eqref{fig:surr075} we show the superior case. Given all data-sets have a fixed $q=0.1$ we show that the bootstrap method performance is correlated to the size of $n$ with the larger sample sizes achieving better outcomes with less iterations. While the growth in performance is positively steep for the first 750 iterations it tends to slow down past that point for the 3 cases ($n = {100, 150, 200}$) after which the algorithm was stopped for reaching maximum iterations.
	
	It is misleading to conclude that larger n is beneficial, this is because $q$ was fixed while $n$ was increasing. The issue relates back to Table \eqref{tab:synth} where in the case of shortest time-series (i.e length 20) with data-sets size going from $N$= {500, 1500, 3000} the respective ARI are 0.68, 0.54, and 0.48. The ARI is negatively correlated with the size of the data-set when time-series length is small. The bootstrap method applied to a similar case with $N$=2000, $n$ = 200, and series length of 20 (See green curve in Fig. \eqref{fig:surr075}) yields an ARI higher than 0.8 after 750 iterations thus demonstrating the usefulness of the method when dealing with large data-sets and short time-series.
	
	\section{Hierarchical Block Correlations} \label{sec:nested}
	
	ALC is efficient at clustering systems of correlated time-series generated using a single factor model when compared to HDBSCAN, but the likelihood model used is one explicitly derived using a single factor. Realistic correlation structures would include more complex time-series models that incorporate multiple factors. In particular, an important related problem in financial time-series clustering is that of clustering correlation matrices with a block hierarchical structure \cite{marti_clustering_2016,tumminello_hierarchically_2007,tumminello_correlation_2010}. Here each block corresponds to a correlation cluster that should be recoverable along with its nested structure. Neither ALC nor HDBSCAN recover the dendrogram, and hence cannot be expected to recover the full nested structure without iterating recursively over each identified cluster until only singletons remain and retaining the configuration paths for these hierarchical iterations. For this reason it is useful to compare the difference between ALC and HDBSCAN on a block hierarchical problem as this will demonstrate more clearly how ALC deals with sub-clusters when they may exists within clusters.     
	
	\begin{figure*}
		\begin{subfigure}{.40\textwidth}
			\includegraphics[width=\textwidth]{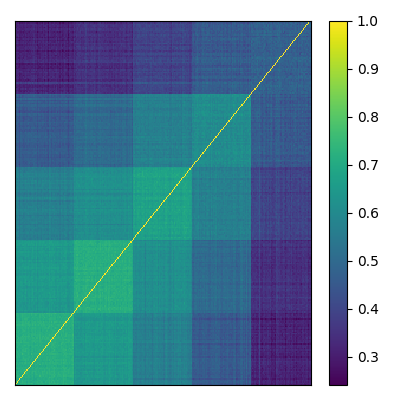}	
			
			\caption{Correlation Matrix of a cluster with 5 hierarchies generated using Eqns. \eqref{eq:nestedfactor1}-\eqref{eq:nestedfactor3}}
			\label{fig:nestedmatrixcluster}
		\end{subfigure}	
		\begin{subfigure}{0.40\textwidth}
			\includegraphics[width=\textwidth]{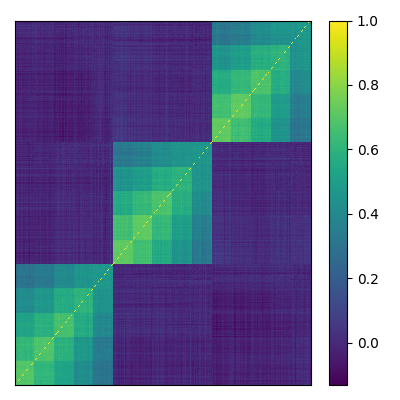}
			\caption{Correlation Matrix of system of 3 clusters with 5 hierarchies generated using Eqns. \eqref{eq:nestedfactor1}-\eqref{eq:nestedfactor3} }
			\label{fig:nestedmatrixsystem}
		\end{subfigure}	
		
		\begin{subfigure}{.45\textwidth}
			\includegraphics[width=\textwidth]{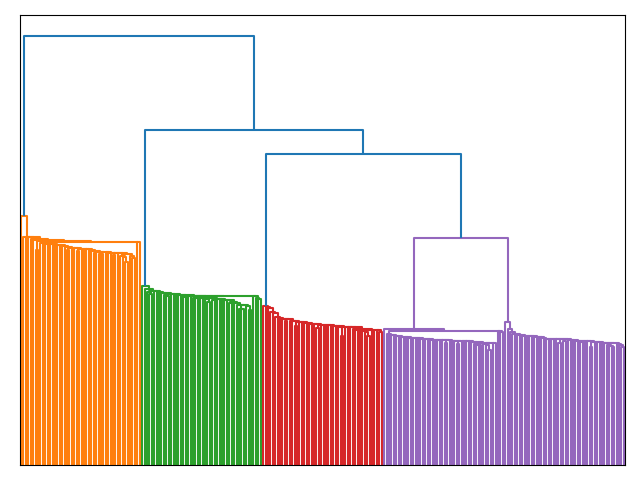}	
			
			\caption{ Dendrogram of a cluster with 5 hierarchies generated using Eqns. \eqref{eq:nestedfactor1}-\eqref{eq:nestedfactor3}}
			\label{fig:dendrogramcluster}
		\end{subfigure}	
		\begin{subfigure}{0.45\textwidth}
			\includegraphics[width=\textwidth]{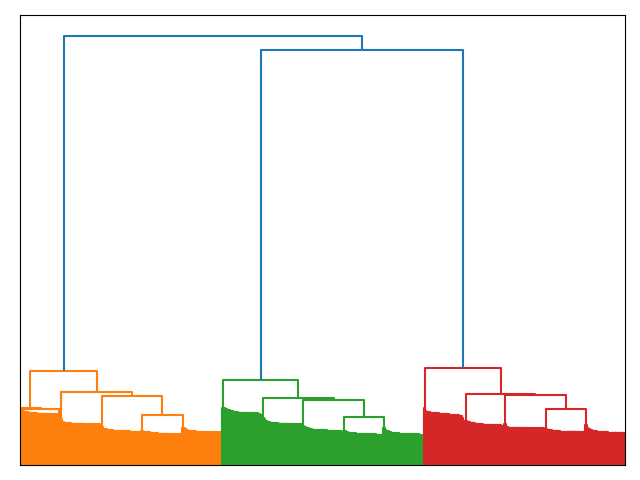}
			\caption{Dendrogram of system of 3 clusters with 5 hierarchies generated using Eqns. \eqref{eq:nestedfactor1}-\eqref{eq:nestedfactor3}}
			\label{fig:dendrogramsystem}
		\end{subfigure}
		\caption{Correlation matrices with nested clusters and their respective dendrograms. On the left is the 5-hierarchy model, and on the right 3 clusters with 5 nested hierarchies.} \label{fig:nested}
	\end{figure*}
	
	Consider $N$ stocks whose price return $x_i$ are influenced by a set of $c$  hierarchical factors $\eta$, and $\epsilon$ the stocks individual effects; here $\eta$ and $\epsilon$ are IID random variables $\sim \mathbf{N}(0,1)$. The correlation matrix can be organized in nested clusters:
	$ h_1 \supset \cdots \supset h_k \supset \cdots \supset h_c$, 
	where $h_1, \cdots, h_c$ represent hierarchies at which factors $\eta_1, \cdots, \eta_c$ are sequentially added to the signal so that along the hierarchical path all stocks are correlated while lower hierarchies inherit all factors from upper ones. The picture is clearer when we consider a model of price return process $x_i$:
	\begin{eqnarray}
	h_1: x_{i} &&= \beta_1 \eta_1 + \alpha_1\epsilon_i, \label{eq:nestedfactor1}\\
	\vdots \notag \\
	h_k: x_i &&= \beta_1 \eta_1 + \cdots+ \beta_k \eta_k + \alpha_k\epsilon_i, \label{eq:nestedfactor2} \\
	\vdots \notag \\
	h_c: x_i &&= \beta_1 \eta_1 + \cdots+ \beta_k \eta_k + \cdots+ \beta_c \eta_c +  \alpha_c\epsilon_i. \label{eq:nestedfactor3}
	\end{eqnarray}
	Here $\alpha_k = 1 - \sum_j^k \beta_j$ where $k$ is the hierarchy index, $\alpha$ modulates the signal to noise ratio and can be arbitrarily set to any value by enforcing a constraint on the sum $\sum_j \beta_j$ to be equal to any value between 0 and 1. The $\beta$s are also chosen such that $\beta_j < \beta_{j+1}$ this ensures that stocks belonging to hierarchy $h_k$ are more correlated to $\eta_k$. There are many ways of parameterizing such nested linear models, and in fact this is just a special case of the more general model of \citet{tumminello_hierarchically_2007}.
	
	Using this model we demonstrate clustering on two hiearchical block correlation examples: a.) a correlation matrix of a cluster of 250 stocks with 5 hierarchies (50 stocks per hierarchy) in Fig. \eqref{fig:nestedmatrixcluster}, and b.) that of a system with 3 clusters estimated on timeseries of 250 observations ({\it i.e.} 1 trading year) for $\alpha = 0.4$ (as shown in Fig. \eqref{fig:nestedmatrixsystem}). Unlike our toy model, real correlation matrices are unlikely to be statically ordered in a way which can reveal their block structure which would trivialize clustering. 
	
	The dendrograms in Figures \eqref{fig:dendrogramcluster} and \eqref{fig:dendrogramsystem} created using the single linkage algorithm reveals the hierarchical structure present in the the correlation matrix in Figures \eqref{fig:nestedmatrixcluster} and \eqref{fig:nestedmatrixsystem}. At first sight looking at the ordered correlation matrices one can distinguish the 5 hierarchies in Figures \eqref{fig:dendrogramcluster} and \eqref{fig:nestedmatrixcluster}. Whereas in system containing multiples of these sets of nested clusters the apparent structure is that of 3 clusters with potential sub-clusters.
	
	\begin{figure*}
		\begin{subfigure}{.45\textwidth}
			\includegraphics[width=\textwidth]{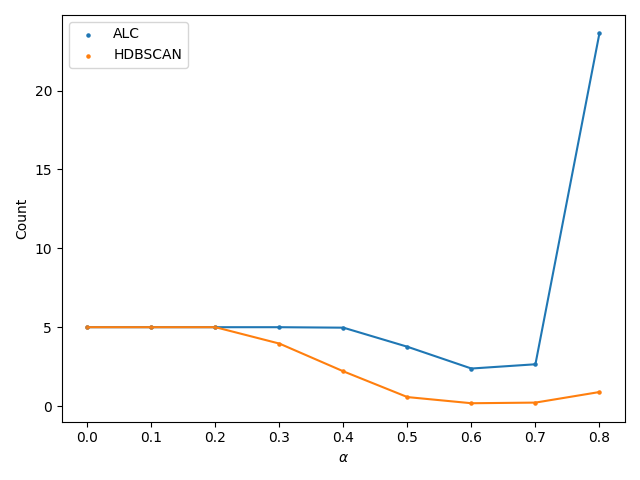}	
			
			\caption{The number of clusters recovered using HDBSCAN and ALC on data in Fig. \eqref{fig:nestedmatrixcluster} as a function cluster specific noise. ALC is more robust to cluster specific noise in the hierarchical correlation model.}
			\label{fig:algosnestedcluster}
		\end{subfigure}	
		\begin{subfigure}{0.45\textwidth}
			\includegraphics[width=\textwidth]{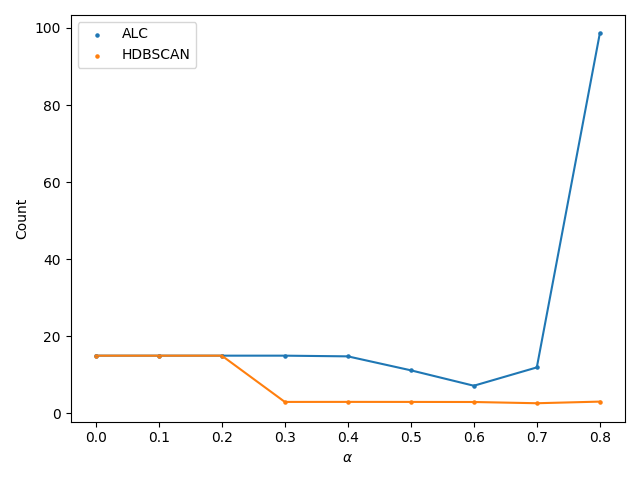}
			\caption{The number of clusters recovered using HDBSCAN and ALC on data in Fig. \eqref{fig:nestedmatrixsystem} as function of cluster specific noise. ALC is more robust to cluster specific noise in the hierarchical correlation model.}
			\label{fig:algosnestedsystem}
		\end{subfigure}	
		
		\caption{The number of clusters as a function of the cluster specific noise level parameter $\alpha$. As $\alpha$ increases the cluster specific noise increases. ALC degrades from $\alpha=0.4$ and at high levels of noise generates many singletons.} \label{fig:nestedclustering}
	\end{figure*}
	
	\begin{figure}
		\includegraphics[width=0.45\textwidth]{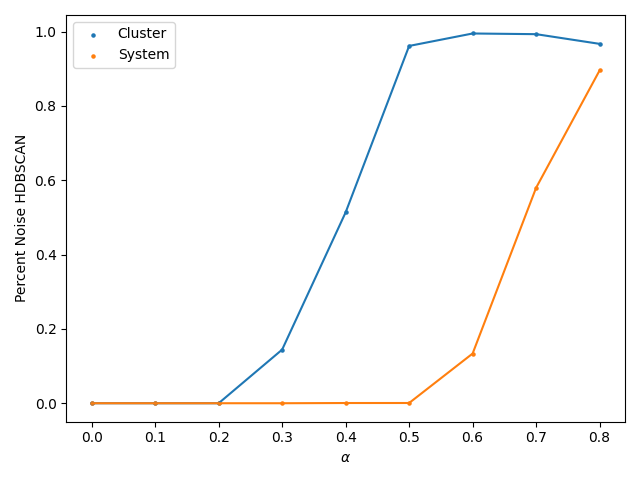}	
		\caption{ The percentage of objects misclassified as noise by HDBSCAN applied to data in ``Cluster'' Fig. \eqref{fig:nestedmatrixcluster} and ``System'' Fig. \eqref{fig:nestedmatrixsystem} }
		\label{fig:hdbscannoise}
	\end{figure}
	
	The key question here is whether or not the solutions resulting from ALC and HDBSCAN contain the subclusters or the large clusters; {\it i.e.} do the results from the data analysis in Fig. \eqref{fig:nestedmatrixcluster} contain 1 or 5 clusters, and those of Fig. \eqref{fig:nestedmatrixsystem}, 3 or 15 clusters? To answer this we have clustered these two datasets for several values of $\alpha$ to investigate how low correlation, and indirectly low cluster density, affect the algorithms' outputs. We show that for low $\alpha$, the timeseries are mostly influenced by the factors, and HDBSCAN and ALC both recover 5 and 15 clusters respectively for the data in Fig. \eqref{fig:nestedmatrixcluster} and \eqref{fig:nestedmatrixsystem}. The effect of noise is stronger for HDBSCAN, where the algorithm starts adding observations to the ``noise'' category for datasets generated with $\alpha > 0.2$ (See Fig. \eqref{fig:hdbscannoise}) for data in Fig. \eqref{fig:nestedmatrixcluster}. While the ALC generated solutions start degrading from $\alpha > 0.4$. Similarly for the data in Fig. \eqref{fig:nestedmatrixsystem} HDBSCAN starts merging the nested clusters into 3 large clusters for $\alpha > 0.2$, and for $\alpha > 0.5$ the solution degrades even further (See Fig. \eqref{fig:hdbscannoise}). Again, the ALC solutions are more resilient and start degrading for $\alpha >0.4$ where nested clusters are merged and the 3 cluster solution is never recovered. 
	
	This marks a distinct difference between ALC and HDBSCAN: ALC seems to prioritize smaller but more densely correlated clusters, whereas HDBSCAN default behavior is to merge clusters even if the resulting cluster is less dense (i.e. noisier). ALC seems to also be 20\% more robust to noise. Finally for $\alpha > 0.7$ ALC solutions are complete disordered partitions.
	
	\section{Performance Improvement} \label{sec:performance}
	
\begin{figure*}
		\begin{subfigure}{.45\textwidth}
			\includegraphics[width=\textwidth]{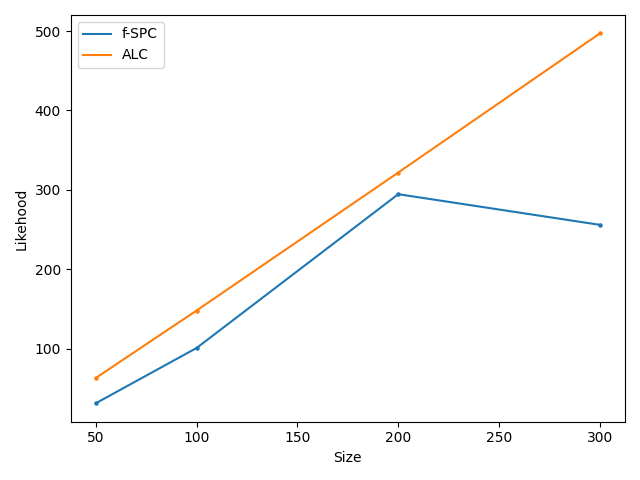}	
			\caption{Comparing cluster quality using the likelihood $L_c$ (See the models in Section \ref{sec:giada_marsili}) of solutions obtained as the data-set size is increased on a log-scale using f-SPC \cite{yelibi_fast_2020} and ALC (Section \ref{sec:alc}). ALC solutions are systematically of higher likelihood than ones obtained using f-SPC.}
			\label{fig:comparison_lc}
		\end{subfigure}	
		\begin{subfigure}{0.46\textwidth}
			\includegraphics[width=\textwidth]{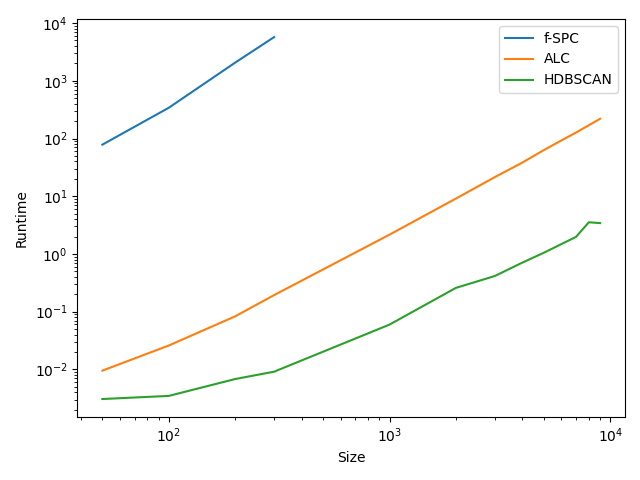}
			\caption{Comparing the algorithm runtime performance in seconds as we increase data-set size on a log-scale for f-SPC \cite{yelibi_fast_2020}, ALC (Section \ref{sec:alc}), and HDBSCAN \cite{mcinnes_hdbscan_2017}. All are roughly quadratic, f-SPC significantly more expensive, while HDBSCAN outperforms ALC.}
			\label{fig:comparison_runtime}
		\end{subfigure}\label{fig:comp}
	\end{figure*}

	The appeal of ALC class of clustering algorithms is its ability to consistently outperform the f-SPC algorithm\cite{yelibi_fast_2020} while being generally robust to noise. Here we clustered data-sets of time-series using the synthetic model described in Appendix \ref{addix:onefactor} with increasing size (N=50,100,200,300) and 10 clusters. Fig. \ref{fig:comparison_lc} shows that ALC recovers clustering solutions of better quality if we use the likelihood as a quality function as was argued in \cite{giada_algorithms_2002}. Furthermore the algorithm runtime is roughly quadratic. This was estimated on synthetic data-sets \footnote{Gaussian mixtures of 10 clusters of sizes ranging from 100 to 10000} $O(N^{1.97})$ and on real data-sets \footnote{ We obtained CRSP US Mutual Funds Net Asset Values (NAV) sampled daily from 1998 to 2020 from which were sampled data-sets of size from 100 to 10000}  $O(N^{2.11})$ thus making it competitive for online learning problems. We compared the algorithm's runtime to the previous f-SPC and HDBSCAN \cite{mcinnes_hdbscan_2017}. Not only f-SPC's solutions have lower likelihood as previously mentioned but require much more computing power to achieve convergence in a significantly longer time\footnote{For ALC and HDBSCAN, we are able to cluster larger data-sets and extend the simulations up to N=10,000. All scripts were executed on a mobile Intel i7-CPU with 4 CPUs. This is stall generation dependent, and for a higher number it should be possible to obtain better results. This illustrates that f-SPC requires non-trivial parameterization, while ALC does not.}.
	
	Figure \ref{fig:comparison_runtime} demonstrates that future optimization algorithms cannot be expected to more expensive than ALC if they provide low quality solutions. Finally, HDBSCAN which we believe the natural alternative to ALC as a hierarchical and density based clustering method, is more optimized than our algorithm, and would thus be currently preferable for extreme large data-sets. However, there is some evidence that suggests that ALC can be considered as a serious alternative to HDBSCAN because under certain circumstances, when data is less dense, less correlated, and noisier, it outperforms in terms of ARI (See Table \ref{tab:synth}). The key point is that ALC looks to be competative both on quality and performance prior to any low level attempts at refining algorithm for optimised runtime performance.
	
	\section{Discussion and Conclusion} \label{sec:discussion}
	
	We have presented an agglomerative algorithm capable of performing the maximization of the Giada-Marsili $L_c$ likelihood (See Eqn \eqref{eq:lc}). In prior work we built and proved a mechanism which maximizes the likelihood locally using Eqn. \eqref{eq:lcmod} instead of Eqn. \eqref{eq:lc} \cite{yelibi_fast_2020}. Here, instead of random moves, we perform a comprehensive search over all the possible combinations and select the optimal move at every iteration. 
	
	The algorithm is considerably faster than Markov Chain Monte Carlo, and Genetic Algorithm based solutions and achieves equal or better maxima. It only requires a correlation matrix as input, and its output is the clustering configuration which reflects the optimal number of clusters of correlated samples. It requires no prior information on the number of clusters. This feature may make the algorithm appropriate for state-detection for online learning in a fast-big-data environment.
	
	We also present a data generation method of simulated correlated time-series based on the Noh ansatz given in Eqn. \eqref{eq:ansatz_1}. Individual time-series are identified by their daily random effect $\epsilon$ while they remain subject to their cluster effect $\eta$, and the strength of that coupling between individual objects and their cluster is expressed by $g_s$. This allowed us to modulate cluster densities, and study its impact on clustering quality and benchmark performance. We showed that even at low intra-cluster coupling values ($g_s \approx 0.05$) the performance is poor, but as $g_s$ increases and clusters densify the algorithm gets better faster than its competing alternative HDBSCAN.
	
	When the signal-to-noise ratio is low it is possible under certain circumstances to mitigate the impact of correlation matrices noise by making use of bootstrapping in conjunction with ALC. This allows for better overall performance, but at the cost of significantly increasing the algorithm runtime. The increased computation related to bootstrap can be carried out in a batch format, or in parallel to the cluster algorithm itself. However, an ability to perform accurate clustering for relatively short timeseries (or equivalently lower dimensional data-sets) is valuable. We note that the exploration of alternate, and perhaps more efficient sampling schemes for the bootstrap step can mitigate this added cost \cite{kim_optimized_2017}. The current Bootstrap method introduced in \eqref{sec:bootstrap} uniformly samples a space of $n$ in $N$ variables which for large $N$ becomes very expensive. Additionally, we demonstrated its use on synthetic data with a known ground truth. On real data, there is a clear need to define a stopping criteria for potential users (i.e. Convergence of the Likelihood $L_c$.).
	
	When we design and cluster datasets using hierarchical factor models, both ALC and HDSCAN recover the underlying nested clusters. ALC is 20\% more robust to noise for the synthetic datasets generated. In fact for large systems of such hierarchical clusters HDBSCAN has a tendency to merge clusters into giant ones. There is no single solution to this kind of clustering problem and it is left to the practitioner to determine whether clustering solutions with large but less dense clusters are more useful than solutions with small but denser clusters. 
	
	Potential further research could be achieved by using this algorithm in an online learning environment. We suspect in the case of financial markets that it may be possible to perform temporal clustering which would allow the analysis of the existing dynamics in financial markets states with increased reliability. Specifically for dynamic cluster analysis of financial markets around recorded extreme events\footnote{Forthcoming work}.
	
	The current version of the algorithm is able to comfortably process data-sets of size up to 10,000 samples \footnote{Here tested on a Intel(R) Core(TM) i7-6700HQ CPU @ 2.60GHz with 32GB of Random Access Memory.} in less than 5 mins -- enough time for a ``cup of coffee''. Crucially, it should also be noted that the Louvain algorithm exists in multiple faster versions than the original implementation use in the current algorithm \cite{traag_faster_2015,fontolan_modularity_2020-1,ozaki_simple_2016} It feasible that one should be able to find further optimizations which will make it possible to cluster massively large dense correlation matrices similarly to what is currently done in the field of Network Science.
	
	The impact on distributional assumptions is also an important area of future work, the Giada-Marsili Likelihood of Section \eqref{sec:giada_marsili} was derived assuming IID Gaussian random variables. Using the same Noz Ansatz it should be possible to arrive at relate model extensions that assume different distributions {\it. e.g.} the Student-t distribution. Concretely, the Gaussian assumptions serves as a baseline for this type of model, but it is known that stock market log returns are at least fat tailed distributed (beside additional stylised facts such as long-memory, volatility clustering and the leverage effect). This motivates alternative modeling with distributions which capture additional stylized fact \cite{gabaix_theory_2003,gopikrishnan_inverse_1998}.
	
	Furthermore, different measures of similarity could be used. The covariance between two random variables is the standard similarity metric used to perform data clustering of financial market assets return series. Alternatives such as the information theoretic Mutual Information have previously been used directly on financial data time-series \cite{fiedor_networks_2014,guo_development_2018}.
	
	Finally we believe it is possible to upgrade ALC with more efficient coding, and superior modifications to the optimization scheme itself similar to recent Louvain's implementations.
	
	\section{Acknowledgements} \label{sec:ackno}
	The authors thank Nic Murphy, Daniele Marinazzo, Gautier Marti and Unarine Singo for discussions and comments.

	\bibliographystyle{apalike}
	\bibliography{bibliography.bib}
	
%	\newpage
	\appendix
	
	% reset the counter for tables, and switch back to arabic
	\setcounter{table}{0}
	\renewcommand{\thetable}{\arabic{table}}
	
	\section{Algorithm: Agglomerative Likelihood Clustering} \label{addix:algo}
	
	Here we provide ALC's pseudocode which: 1.) generates clustering candidates, 2.) evaluates
	the likelihood of the candidate configurations $L_c$ \cite{giada_data_2001} using Eqn. \eqref{eq:lcmod}, and then 3.) selects the best candidates and discards the others.
	
	\begin{table}[H]
		\caption{\label{alg:alc}}
		\begin{algorithm}[H]
			\caption{Pseudo-code for a ALC implementation (Sec. \eqref{sec:alc}), and ``alc.py" in \cite{yelibi_agglomerative_2020}}
			\begin{algorithmic}[1]
				\STATE {INPUT: Correlation Matrix, OUTPUT: Tracker }
				
				\STATE{Produce population of size N}
				\WHILE{	N $\geq$ 2}
				\STATE {Pick label i in population}			
				\FOR { j $\neq$ i in population  }
				\STATE{create label k by merging labels i and j}
				\STATE{ Pick k by optimizing for $\Delta L_c$}
				\STATE{add k, remove i and j from population. }
				\ENDFOR
				\STATE { Stop if $\Delta L_c \leq 0$ and output Tracker }

				\FOR { j in population }
				\STATE {store correlation $C_{kj}$ in correlation matrix }
				\STATE {update cluster composition in Tracker}
				\ENDFOR
				\ENDWHILE
				
			\end{algorithmic}
		\end{algorithm}
	\end{table}
	
	\section{Data Generative Process: Synthetic Correlated Time-Series} \label{addix:onefactor}
	
	The Noh Ansatz \cite{noh_model_2000,giada_data_2001} offers a powerful stochastic process model for clustering purposes. We make use of the following equation (equivalent to Eqn. \eqref{eq:ansatz_1} ) as a way of generating correlated time-series:
	\begin{equation} \label{eq:ansatz_2} \xi_i(d) = \frac{\sqrt{g_{s_i}}\eta_{s_i}(d) + \epsilon_i(d) }{\sqrt{1+g_{s_i}} } \end{equation}
	
	The process is described in Table \eqref{tab:gen_timeseries}. In Fig. \eqref{fig:correlated_timeseries} We plotted 3000 simulated time-series of $D \in  (20, 60, 250)$ with 10 clusters each of size 300. Time-series visualizations lack interpretability especially given apparent noise and chaos. We use additional tools such as Minimum Spanning Trees (MST) and dimensionality reduction methods such as UMAP to provide better visualization. Shown in Fig. \eqref{fig:mst_gm} and \eqref{fig:mst_est} are respectively the MSTs for the true correlation matrix, and the estimated of the time-series returns in Fig. \eqref{fig:correlated_timeseries}. Because MSTs rely on nearest neighbors to build the graph, here the input is $d_{ij} = 1 - \rho_{ij}$. The MSTs are clearly capable of dissociating clusters, and one could, in this simple case, discern all 10 clusters without much prior information. A minor remark is in the topology of the MST which differs from the true to the estimated correlation matrix: estimating correlations from data introduces noise which is reflected in its topology \footnote{ Noise due to the estimation of the correlation matrix are typically cleaned to a certain extent using Random Matrix Theory methods \cite{bun_cleaning_2017,wilcox_analysis_2007}}.

	\begin{table}[H]
		\noindent\fbox{\parbox{.45\textwidth}{
				\begin{enumerate}[label=\eqref{tab:gen_timeseries}\arabic*]
					\item  {\bf Cluster Number}: Define values for number of cluster $C$, and size of clusters $s$ and obtain $N = s*C$ the number of time-series in the data-set. Pick a time-series length $D$
					\item  {\bf Spin-Labels}: Create a list of array of spin-labels with the $C$ labels
					\item {\bf Random Effects}: Create a CxD array $\eta \sim \mathcal{N}(0,\,1)$, and another NxD array $\epsilon \sim \mathcal{N}(0,\,1)$. $\eta~\mbox{and}~\epsilon$ respectively capture the daily cluster and individual object random effects.
					\item  {\bf Fix Intra-cluster Binding Strength}: Pick a value for $g_s$ per cluster: it is not needed that $g_s$ be identical for every clusters. As a matter of fact, real noisy systems will display various $g_s$ values. However fixing $g_s$ simplifies the process and increases interpretability.
					\item  {\bf Compute Returns}: Create a NxD array $\xi$ and compute the daily returns using Eqn. \eqref{eq:ansatz_2} by looping over the clusters, and the time-series within the clusters.
				\end{enumerate}
		}}
		\caption{ \label{tab:gen_timeseries} Implementation of Noh Ansatz model of correlated time-series (See code at \cite{yelibi_one_2020} ) }
	\end{table}
	
	\begin{figure*}
		\includegraphics[width=\textwidth]{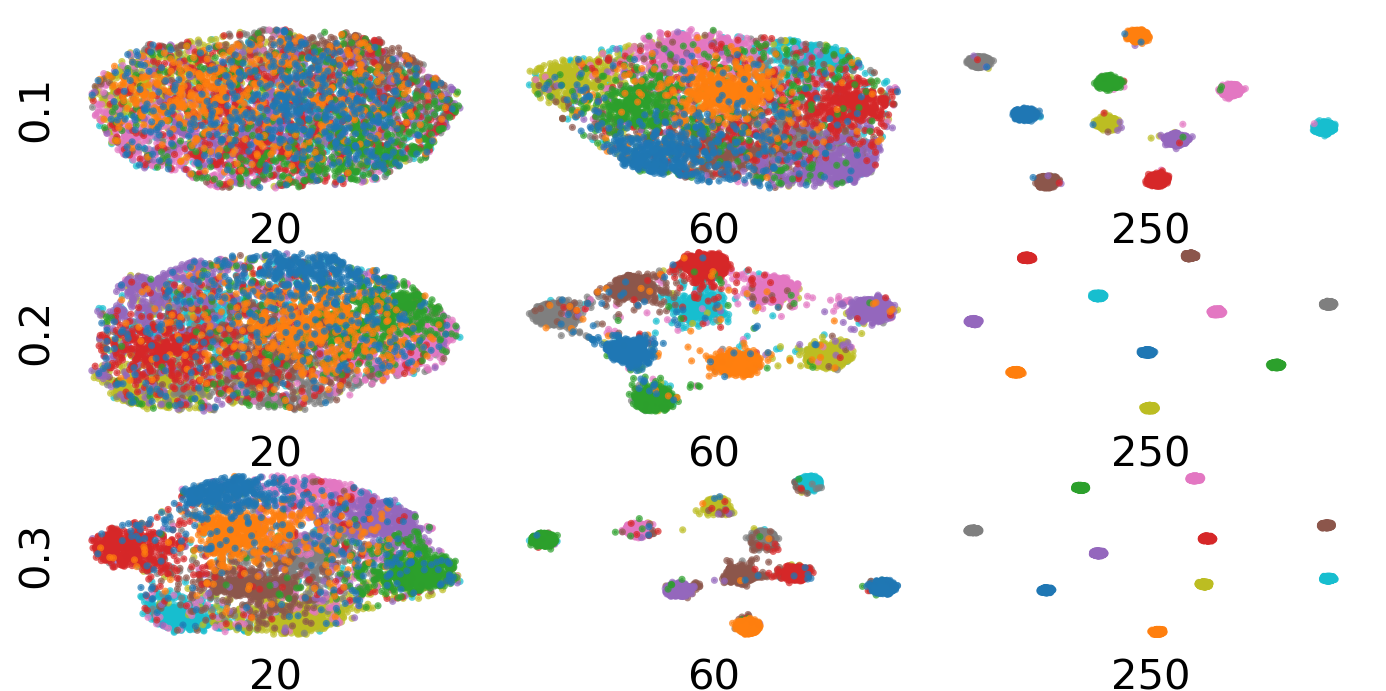}
		\caption{From top to bottom $g_s$ is varied from 0.1 to 0.3 (See model in Appendix \eqref{addix:onefactor}	) and from left to right, time-series length from 20 to 250. The result is the UMAP \cite{mcinnes_umap_2018} 2D embed of 3000 correlated time-series sorted within 10 clusters.}\label{fig:correlated_umap}
	\end{figure*}

	Using UMAP we plot the same data in Fig. \eqref{fig:correlated_timeseries} but projected on a 2D manifold in Fig. \eqref{fig:correlated_umap} for visual purposes only. We expect UMAP to capture the local structure of the data: From left to right the time-series lengths are 20 to 250, while from top to bottom $g_s$ values range from 0.1 to 0.3. The noise level is captured by $g_s$: clusters with $g_s \rightarrow 0$ are diffuse and close to each other whereas those with $g_s \rightarrow 1$ show increasingly high density. Similarly short time-series induce spurious correlations and noise within clusters. The data-sets remain modular and very diffuse but as time-series length increases they spread out and thus should become easier to separate by clustering algorithms.
	
\end{document}